\documentclass{aa}
\usepackage{natbib}   
\bibpunct{(}{)}{;}{a}{}{,}    

\usepackage{amsmath}
\usepackage{graphicx}
\usepackage{txfonts}

\newcommand\Sextractor{\texttt{SExtractor}}
\newcommand\Hyperz{\texttt{hyperz}}
\newcommand\cB{{\it b}} 
\newcommand\cF{{\it f}}
\newcommand\cD{{\it d}}
\newcommand\cP{{\it p}}

\newcommand\PLOS{190}    
\newcommand\NTF{2\,064}  
\newcommand\NSP{30\,444} 
                         
\newcommand\Rabs{M_{r}}
\newcommand\kmps{\mathrm{km\,s^{-1}}}
 \newcommand\Mpc{\mathrm{Mpc}}
 \newcommand\der{\mathrm{d}}
  \newcommand\zl{z_\mathrm{l}}
  \newcommand\zs{z_\mathrm{s}}
  \newcommand\Ds{D_\mathrm{os}}
  
 \newcommand\Dls{D_\mathrm{ls}}
\newcommand\rcut{r_\mathrm{cut}}
\newcommand\rarc{r_\mathrm{arc}}
  \newcommand\re{r_\mathrm{E}}
\newcommand\reTF{r_\mathrm{E(TF)}}

\newcommand\finbf[1]{{#1}}

\begin{document}

\title{A search for edge-on galaxy lenses in the CFHT Legacy Survey
  \thanks{Based on observations obtained with MegaPrime/MegaCam, a joint
    project of CFHT and CEA/DAPNIA, at the Canada-France-Hawaii Telescope (CFHT)
    which is operated by the National Research Council (NRC) of Canada,
    the Institut National des Sciences de l'Univers of the Centre National
    de la Recherche Scientifique (CNRS) of France, and the University of Hawaii.
    This work is based in part on data products produced at TERAPIX
    and the Canadian Astronomy Data Centre as part of the Canada-France-Hawaii
    Telescope Legacy Survey, a collaborative project of NRC and CNRS.
  }
}

\author{
  J.F. Sygnet\inst{1,2}
  \and
  H. Tu\inst{3,1,2}
  \and
  B. Fort\inst{1,2}
  \and
  R. Gavazzi\inst{1,2}
}

\institute{ 
CNRS, UMR7095, Institut d'Astrophysique de Paris, F-75014 Paris, France 
\\ \email{[sygnet;fort;gavazzi]@iap.fr}  
\and
UPMC Univ Paris 06, UMR7095, Institut d'Astrophysique de Paris, F-75014 Paris, France
\and
Physics Department \& Shanghai Key Lab for Astrophysics, 
Shanghai Normal University, Shanghai 200234, China 
\\ \email{TuHong@shnu.edu.cn}
}

\date{Received 28 December 2009 /  Accepted 12 April 2010}

\abstract 
   {  
The new generation of wide field optical imaging like the 
Canada France Hawaii Telescope Legacy Survey (CFHTLS) enables discoveries of 
all types of gravitational lenses present in the sky.
The Strong Lensing Legacy Survey (SL2S) project has  started
an inventory, respectively for clusters or groups of galaxies lenses,
and for Einstein rings around distant massive ellipticals.  
   }
   {  
Here we attempt to extend this inventory by finding lensing events produced by 
massive edge-on disk galaxies which remains a poorly documented class of lenses.
   }
   {  
We implement and test an automated search procedure of edge-on galaxy lenses 
in the CFHTLS Wide fields with  magnitude $18<i<21$, inclination angle lower 
than $25\degr$ and having a photometric redshift determination. 
The procedure estimates the lensing convergence of each galaxy from the
Tully-Fisher law and selects only the few candidates which exhibit a possible 
nearby arc configuration at a radius compatible with this convergence 
($\rarc \lesssim 2 \re $).
The efficiency of the procedure is tested after a visual examination of the 
whole initial sample of \NSP\ individual edge-on disks. 
   } 
   {  
We calculate the surface density of edge-on lenses possibly detected in a survey
for a given seeing. 
We deduce that this theoretical number is about 10 for the CFHTLS Wide, 
a number \finbf{in broad agreement} with the 2 good candidates detected here.
We show that the Tully-Fisher selection method is very efficient at 
finding valuable candidates, albeit its accuracy depends on the 
quality of the photometric redshift of the lenses.
Eventually, we argue that future surveys 
will detect at least a hundred of such lens candidates.    
   } 
   {  
   }
 
 
\maketitle

\section{Introduction} 

The $\Lambda$-CDM cosmological paradigm has been highly successful at explaining most
of the large scale properties of the Universe. Nevertheless it is still to be shown
that this model is able to explain its small scale properties. 
Among these, the formation and dark matter content of disk galaxies, and in 
particular the rotation curves of disks played a key role at giving evidence for the 
presence of dark matter on small scales \citep[e.g.~][]{Bosma1978, Rubin1980, Persic1996, deBlok2008,Donato2009}. 
Despite tremendous observational and modeling efforts over the last three decades, 
\finbf{advances} are still plagued by our ability to disentangle both
the contributions of baryonic disk and bulges and the contribution of dark matter.

The hypothesis of the maximum bulge or disk \citep{vanAlbada86, Persic1996, Dutton2005},
is able to fix the mass-to-light ratio of the stellar populations and 
therefore to put constraints on the dark matter content of galaxies or alternatively 
on inferences about modifications of the gravitation law 
\citep[e.g.~][]{deBlok1997, McGaugh2004,deBlok2008,Donato2009}.
In order to get further insight onto the relative contribution of different species 
(stars in disk, stars in bulge and dark matter) one needs to get additional 
constraints to break the degeneracies \citep[e.g.~][]{Kranz2003,deJong2007,Ibata2007}. 

Another particularly promising method resides in studying the dynamical properties of strong 
lensing galaxies. Until recently, essentially due to the scarcity of background QSOs, 
only very few lensing disk galaxies with a low enough redshift, suitable for 
kinematical observations, were known. 
For instance, one can mention the early studies of 
the Einstein Cross Q2237-0305 \citep{Trott2002,Trott2009, vandeVen2008}, 
B1600+434 \citep{Jaunsen1997, Koopmans1998, Maller2000}, J2004-1349 
\citep{Winn2003}, CXOCY J220132.8-320144 \citep{Castander2006} 
or Q0045-3337 \citep{Chieregato2007}.

Great \finbf{advances} have become recently possible thanks to the advent of 
large high-quality imaging and spectroscopic datasets. 
Automated or visual inspection of high resolution imaging from 
Hubble Space telescope (HST) \citep[e.g.~][]{Marshall2009, Newton2009, Faure2008, Moustakas2007, Covone2009} or from the ground \nocite{Gavazzi2010} 
\citep[e.g.~][Gavazzi et al. 2010]{Cabanac2007-SL2S,Kubo2008} 
have just started providing us with many candidate strong galaxy-galaxy lensing events. 
For spectroscopy, an automated search of superimposed spectra at different redshifts 
into a given fiber followed-up by high resolution HST imaging turns out to be very 
efficient on large spectroscopic datasets like the Sloan Digital Sky Survey (SDSS) \citep{Bolton2004, Bolton2006, Bolton2008-ACS5, Auger2009, Willis2005, Willis2006}.
These latter studies are limited to relatively low redshifts
($z\sim 0.2$) and most of them are focused on lensing by massive elliptical galaxies
with large splitting angle.

Conversely, late-type galaxies, while more numerous 
\citep[e.g.~][]{Bartelmann2000, Kochanek2006-SAASFE, Chae2010}, 
have a smaller Einstein radius and are thus more difficult to identify.
For instance the recently found edge-on disk galaxy OAC--GL J1223-1239 
\citep{Covone2009}, with $\re \sim 0\farcs43$, could only be found 
with HST imaging.
A first automated spectroscopic search for disk galaxies lenses in 
the SDSS database was done by \cite{Feron2009}:
they found 8 candidates among 40 000 disk galaxies, two of them being 
already confirmed as genuine lenses.

Despite these efforts, disk galaxies remain a poorly documented 
class of lenses, particularly for edge-on lensing galaxies. 
This situation calls for an improvement.
We need more edge-on disk lenses with a low inclination to maximize 
the success of dynamical studies and simplify the recognition of 
lensing/lensed structures in imaging data.

In this paper we investigate if the deep multiband sub-arcsecond imaging data
of the Canada France Hawaii Telescope Legacy Survey (CFHTLS) offers a good opportunity 
to find edge-on disk lenses beyond the redshift range accessible by the SDSS.

The paper is organized as follows. 
Section \ref{sec:expected} presents a calculation of the number of lenses 
with an $i$ magnitude  $18 < i < 21$ and an inclination lower than 
$25\degr$ which should be detected for a given seeing in an imaging survey. 
Section \ref{sec:search} explains the procedure that we have followed to 
extract edge-on lens candidates from the CFHTLS Wide survey by estimating 
an Einstein radius from the Tully-Fisher (TF) relation \citep{Tully-Fisher1977} 
and comparing it with the arc radius possibly  detected around the galaxy. 
We also cross-check the validity of this automated procedure 
by a visual inspection of all the edge-on galaxies present in the survey. 
In Sect.~\ref{sec:discussion} we discuss our results with an overview on what could 
be done with future all-sky imaging surveys.

Throughout this paper all magnitudes are expressed in the AB system,
and we assume the concordance cosmological model with  $h = 0.7$, 
$\Omega_{\rm m}=0.3$, and $\Omega_\Lambda=0.7$.

\section{Forecasting the frequency of edge-on disk lenses} \label{sec:expected}

\finbf{We attempt here} to predict the number of lenses
that could be observed within an imaging survey of properties similar
to the CFHTLS Wide.

\subsection{Cross-sections and optical depths} \label{ssec:expected1}
We estimate lensing optical depths by starting from generic assumptions about mass
distributions \citep[see e.g.][]{Turner1984, SEF92, Marshall2005, Kochanek2006-SAASFE}.
In particular, we assume that all lenses are Singular Isothermal Spheres 
(SIS) parameterized by their velocity dispersion $\sigma$. 
They are also assumed to be sparse enough that the probability of multiple 
lensing is negligible, so that one can sum up individual lensing 
cross-sections over the probed volume. 
Arguably, circular symmetry might not be a good description of edge-on disk galaxies.
\citet{Bartelmann1998} showed that, averaging over all possible inclination 
angles, the effect of the disk is null for the overall cross-section. 
It is however found that the efficiency of edge-on lenses is a bit boosted 
so that a calculation of the total SIS cross-section and the same calculation
restricted to inclinations $\alpha_i \le 25\degr$  (corresponding to a third
of all possible orientations) consistent with edge-on disks will bracket the 
right number.
So we define a corrective term $1/3<\eta<1$ that will multiply the total lensing
cross-section of disky galaxies approximated as SIS.

Ideally, the cross-section $X$ for multiple imaging by a SIS mass distribution 
at redshift $\zl$ of a source at $\zs$ is given by the solid angle 
(in steradians) subtended by the Einstein radius:
\begin{equation}
  X= \pi \re^2 = \pi \left[ 4\pi \left( \frac{\sigma}{c}\right)^2 \frac{\Dls}{\Ds}\right]^2\,.
\end{equation}
However, due to observational limitations, one cannot detect all the lenses 
of a given $\re$, the main limitation being seeing and blending with the 
light possibly coming from the foreground deflector. 
In addition, for a SIS profile, two images (arcs) are produced at radii 
$\theta_1$ and $\theta_2$ from the optical axis (with the convention 
$\theta_1>\theta_2>0$) so that the radius of the source is
$\beta=(\theta_1-\theta_2)/2$ and its Einstein radius is 
$\re=(\theta_1+\theta_2)/2$.
The detectability of a lens implies that the outermost arc 
$\theta_1\equiv\rarc > \rcut$, with $\rcut$ a limiting radius that we 
set to the seeing FWHM $\rcut=0\farcs8$. 
The other condition for multiple imaging, $\beta<\re$, complements the 
constraints for the actual lensing cross-section.
The full calculation yields
\begin{equation}
X^\prime = 
  \begin{cases}
  0                           & \text{if}\quad  \re <  \rcut/2              \\
  \pi\,(2\re-\rcut) \, \rcut  & \text{if}\quad \rcut/2 \leq \re \leq \rcut  \\
  \pi\,\re^2                  & \text{if}\quad \rcut   <    \re             \\
  \end{cases}
\end{equation}

The optical depth $\tau$ for strong lensing of a source at $\zs$ can be 
written as the volume integral of the above cross-section integrated over 
the distribution of the velocity dispersion distribution function of lenses
$\phi(\sigma)$ that is assumed to be constant with cosmic time (at least 
back to redshift $\sim 1$ beyond which lensing efficiency quickly falls off).
\begin{equation}
\label{eq:tau}
  \tau(\zs) = \int_0^{\zs} \der \zl\: f_k(\chi)^2 \frac{\der \chi}{\der \zl} 
              \int_0^\infty \der \sigma\: 
              K(\zl,\sigma) X^\prime(\sigma,\zl,\zs) \phi(\sigma) \,.
\end{equation}

In this equation we have introduced a 
selection function $K(\zl,\sigma)$ of potential lenses
that will depend on observational settings and lens finding strategy. 
Notice that a purely source oriented approach, typical of radio surveys 
\citep{Browne2003, Winn2003}, will have $K(\zl,\sigma)=1$.
Conversely a lens oriented survey like SLACS \citep{Bolton2006} will be limited 
in redshift by the depth of SDSS spectroscopy and will pick the most massive
 $\sigma\sim 240\,\kmps$ galaxies from the start. 
Since we assume a flat cosmology we can use comoving distances
$f_k(\chi)=\chi$ in Eq.~\eqref{eq:tau}.
We use the velocity dispersion of \citet{Chae2010} for SDSS late-type 
galaxies which reads:
\begin{equation}
\phi(\sigma)  =  \phi^*  \sigma^{-1}  \left( \frac{\sigma}{\sigma^*}\right)^\alpha 
  \exp\left[-\left(\frac{\sigma}{\sigma^*}\right)^\beta\right]
  \frac{\beta}{\Gamma (\alpha / \beta)} \,,
\end{equation}
with 
$\alpha=0.69, ~\beta=2.10, ~\phi^*=0.066 \, h^3 \Mpc^{-3}$ and
$\sigma^*=91.5 \, \kmps$.

\subsection{Selection function of lenses}
As already anticipated, observational limitations will lead to a non-unity 
selection function $K(\zl,\sigma)$. 
Obviously one cannot select directly upon $\zl$ or $\sigma$, but more realistically 
the apparent magnitude is used.
Therefore we assume that a combination of photometric redshifts (with negligible
 errors at this stage) and a direct relation between luminosity and velocity 
dispersion will approximate most of the observational selections effects. 
More precisely, assuming that lens galaxies are isothermal spheres allows 
us to relate the 
velocity dispersion to the rotation velocity $ \sigma = V_c/\sqrt{2}$.
We then use the latest normalization of the Tully-Fischer relation in the R band 
from \citet{Pizagno2007} and neglect the intrinsic scatter about 
the mean relation, giving:

\begin{equation}
\log V_c  = -0.135 \left(  \Rabs - 5 \log_{10} h + 20.332 \right) + 2.210
\end{equation}
In addition, we shift apparent $i$ band magnitudes back to the rest frame 
$r$ band using a redshift dependent k-correction so that \\
$\Rabs= i -25 - 5 \log_{10} ({D_{\rm lum}}/{h^{-1}\Mpc}) - k_{i \to r} .$
A polynomial fit to the photometric change $k_{i \to r}$ yields:

\begin{equation}
  k_{i \to  r}(z) = -0.36 + 0.65 z -0.92 z^2 + z^3 +0.06 z^4 
\end{equation}
which is obtained by redshifting template of Sbc galaxies from \citet{CWW80} 
using \Hyperz\ facility \citep{Bozonella2000-HYPERZ}. 

Hence a given selection in magnitude $i_{\rm min}<i<i_{\rm max}$ can readily 
be cast into a redshift-dependent selection in velocity dispersion: 
$K(\zl,\sigma) =  \Theta(  \sigma -\sigma_{\rm min}(\zl) ) \Theta 
( \sigma_{\rm max}(\zl) - \sigma)$, 
where $\Theta$ is the Heavyside step function.

\subsection{Population of background sources}
Given the lensing optical depth of Eq.~\eqref{eq:tau} and a redshift 
distribution of sources, the number of gravitational lenses can be written as: 
\begin{equation}
	N =  \eta \, A \, n_0 \int_0^\infty \der \zs \:  p(\zs) \, \tau(\zs) \;,
\end{equation}
where $A$ is the surveyed area and $\eta$ the term
accounting for
the boost of the disk introduced in \S~\ref{ssec:expected1}. 
In principle the redshift distribution inferred from photometric redshifts by 
\citet{couponZPHOT} for the CFHTLS Wide survey should be most suitable for 
our analysis. 
However, magnification bias, which can create a gain of more than 
one magnitude in depth, makes better suited the redshift distribution of 
the HST COSMOS survey down to $i<26$, which is closer to the population of faint 
blue objects we aim at detecting behind foreground edge-on disks. 
Therefore we use the results of \citet{leauthaud2007} from COSMOS that 
can be approximated by the expression:
\begin{equation}
\label{eq:ndez}
   \frac{{\der} n (\zs)}{{\der}\zs} = \frac{1}{z_0\Gamma(a)}e^{-\zs/z_0}
   (\zs/z_0)^{a-1}
\end{equation}
with $z_0=0.345$ and $a=3.89$ \citep{Gavazzi2007-SLACS4}. 
In addition, at these limiting $i<26$ magnitudes, one can achieve a 
surface density of sources of about $n_0 = 300\,000\;{\rm deg}^{-2}$.

\subsection{Results}
\begin{figure}
  \includegraphics[width=9cm]{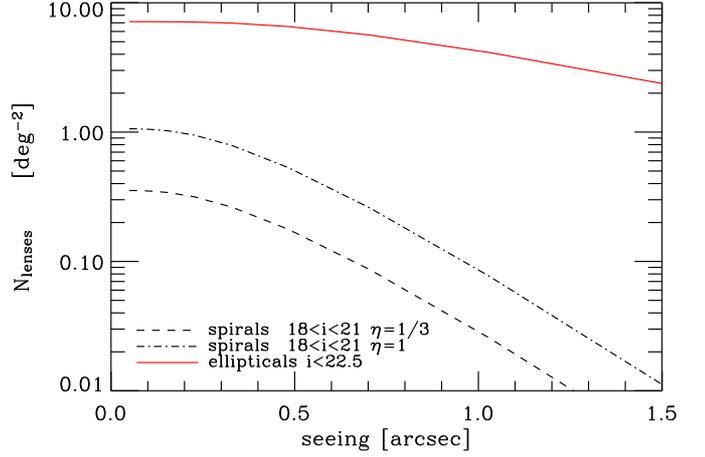}
  \caption{Predicted number of gravitational lenses per square degree as 
   a function of image quality. 
   Imaging survey specifications in terms of depth are those of the CFHTLS Wide. 
   The dashed and dot-dashed black lines bracket the number of lensing 
   edge-on disks depending on the edge-on disk boost factor $\eta$. 
   The upper \finbf{solid} red curve is for lensing by bright $i<22.5$ ellipticals.}
   \label{fig:nlens}
\end{figure}

For the values considered above, we plot in Fig.~\ref{fig:nlens} the predicted 
number of edge-on disks as a function of the cut radius which is identified 
to the seeing.  
For the CFHTLS Wide image quality (seeing $\simeq 0\farcs8$ in 
$g$ band) and depth we typically get about 0.07 -- 0.30 detectable lenses per 
square degree depending on the edge-on disk boost factor $\eta$. 
Therefore for the 124 deg$^2$ imaged in five $u^*griz$ bands, one might discover
between 8 and 37 lenses assuming all our hypotheses (SIS density profile, no 
scatter in the TF relation nor errors with photometric redshifts...) are correct.

To assess our calculations we also made predictions for the number of lensing
elliptical galaxies assuming a perfect relation between magnitude and velocity 
dispersion as given by the Faber-Jackson relation whose normalization is taken 
from \citet{Oguri2006}. 
Down to a limiting AB magnitude $i<22.5$ for the deflector, a similar
calculation based on the velocity dispersion function for ellipticals also 
given by \citet{Chae2010} yields the red curve in Fig.~\ref{fig:nlens}. 
It predicts 10 to 100 more lenses with a substantially lower dependence
on seeing below $1\farcs5$. 
This might be explained by the generally larger Einstein radius (due to greater
velocity dispersions) of ellipticals for which the exponential fall-off of 
the distribution functions drops beyond $1\farcs5$. Worse seeing will also make 
the detectability of arcs and counterimages much more challenging beyond $0\farcs8$.
In addition, the light from the deflector, no matter the seeing size, will 
prevent many small scale detections. 
For comparison, \citet{Marshall2005} predicted about 50 lenses that could 
be detected by SNAP (two magnitudes deeper with FWHM$\sim 0\farcs12$). 
Ten or so systems would remain by degrading depth and image quality to 
our survey specifications and by restricting ourselves to bright $i<22.5$
deflectors. This is consistent with our calculations. 
This is also in broad agreement with COSMOS observations of strong 
lensing luminous 
$\Rabs < -20$ ellipticals \citep{Faure2008}.

\section{Seeking edge-on disk lenses within the CFHTLS Wide}
\label{sec:search}

From the calculations above, we can undertake a comprehensive search of  
lenses in the CFHTLS Wide survey. 
The T0005 data release \citep[for a detailed description, see][]{Mellier2008}
spans 171 fields of one square degree each
for which images in at least the three $g$, $r$, and $i$ bands are obtained.
But, in order to calculate the photometric $\zl$ of the possible lenses,
we restrict ourselves to the 124 fields for which the five  $u^*$, $g$, $r$, 
$i$, and $z$ bands are present.
The typical 80\% completeness limit magnitude for point sources is
25.4, 25.4, 24.6, 24.5, and 23.6 respectively. Image quality is also 
very good with typical seeing FWHM of $0\farcs90,\; 0\farcs85,\;  
0\farcs75,\; 0\farcs72,$ and $0\farcs71$ respectively.

\subsection{Photometric preselection of bright edge-on galaxies}
\label{ssec:photometry}
We start from a catalogue of 28 million extended objects
detected with \Sextractor\ \citep{Bertin1996-SEXTRACTOR}. 
We first select 1.02 million objects within the magnitude range $ 18 < i < 21 $
as the initial data base. The bright $ i < 18 $ low redshift galaxies are 
excluded because their angular size is too large as compared to their 
expected Einstein radius. 
Objects fainter than $ i = 21 $ are also discarded because of the low
signal-to-noise and the difficulty of future spectroscopic follow-up.
The few already known spiral galaxies producing multiply-imaged QSOs 
happened to be in this magnitude range 
\citep{Koopmans1998,Maller2000,Castander2006,Chieregato2007}. 

Edge-on disk galaxies are then preselected as highly elongated objects with 
projected ellipticity $ 0.7 <  e\equiv  (a^2-b^2)/(a^2+b^2) < 0.92 $. 
The lower limit on the ellipticity is chosen because it removes most of 
the ellipticals \citep{Park2007} while the upper limit $e<0.92$ stands for 
rejecting most ($\sim\,$90\%) of spurious objects like diffraction 
patterns of bright stars. 
This ellipticity cut yields \NSP\ bright edge-on galaxies.

\subsection{The Tully-Fischer relation as a proxy for Einstein radius}
\label{ssec:TF}
The multiband ($u^*griz$) photometry of the survey allows the computation 
of photometric redshifts \citep{couponZPHOT}. 
Along with apparent magnitudes, this additional information enables 
estimates of absolute magnitudes that can readily be converted into an 
estimate of their rotation velocity $V_c$ using the latest calibrations 
of the Tully-Fischer relation \citep{Pizagno2007}. 
Assuming the lensing galaxy can be approximated by a circularly symmetric
isothermal mass distribution, the velocity dispersion is
$\sigma = V_c / \sqrt{2} $. 
Using a typical source redshift $1<\zs<3$ and taking into account 
a possible error $\Delta M_r=-0.37$ on the absolute magnitude 
(corresponding to the $1\sigma$-dispersion of the TF law), we can predict 
an Einstein radius $\reTF$ that is then compared to the seeing size.
More specifically, given that the maximum distance of the furthest of the two 
multiple images of a source is twice its Einstein radius and assuming that this 
latter image could not be detected if embedded within the seeing disk, 
we have selected foreground objects by requiring they satisfy 
the TF cut: $ 2 \reTF \ge 0\farcs8$. 

Applying this last automated cut on the catalogue provides us with 
\NTF~massive edge-on disk galaxies with the greatest chance of being lenses.

\subsection{Subsequent visual classification} \label{ssec:visual}
At this stage, we were unable to avoid visual inspection of color cutout 
images of the above \NTF~objects.
The first reason for that is that the small size of Einstein radii and therefore 
distance of multiple images to the central deflector is comparable to both the 
deflector size and seeing disk. 
Hence \Sextractor\ will fail at deblending arcs and deflector.

Consequently we undertook a systematic inspection of these \NTF~objects and 
decided to further consider objects that fulfill the following criteria.

\begin{figure}
   \includegraphics[width=4.cm]{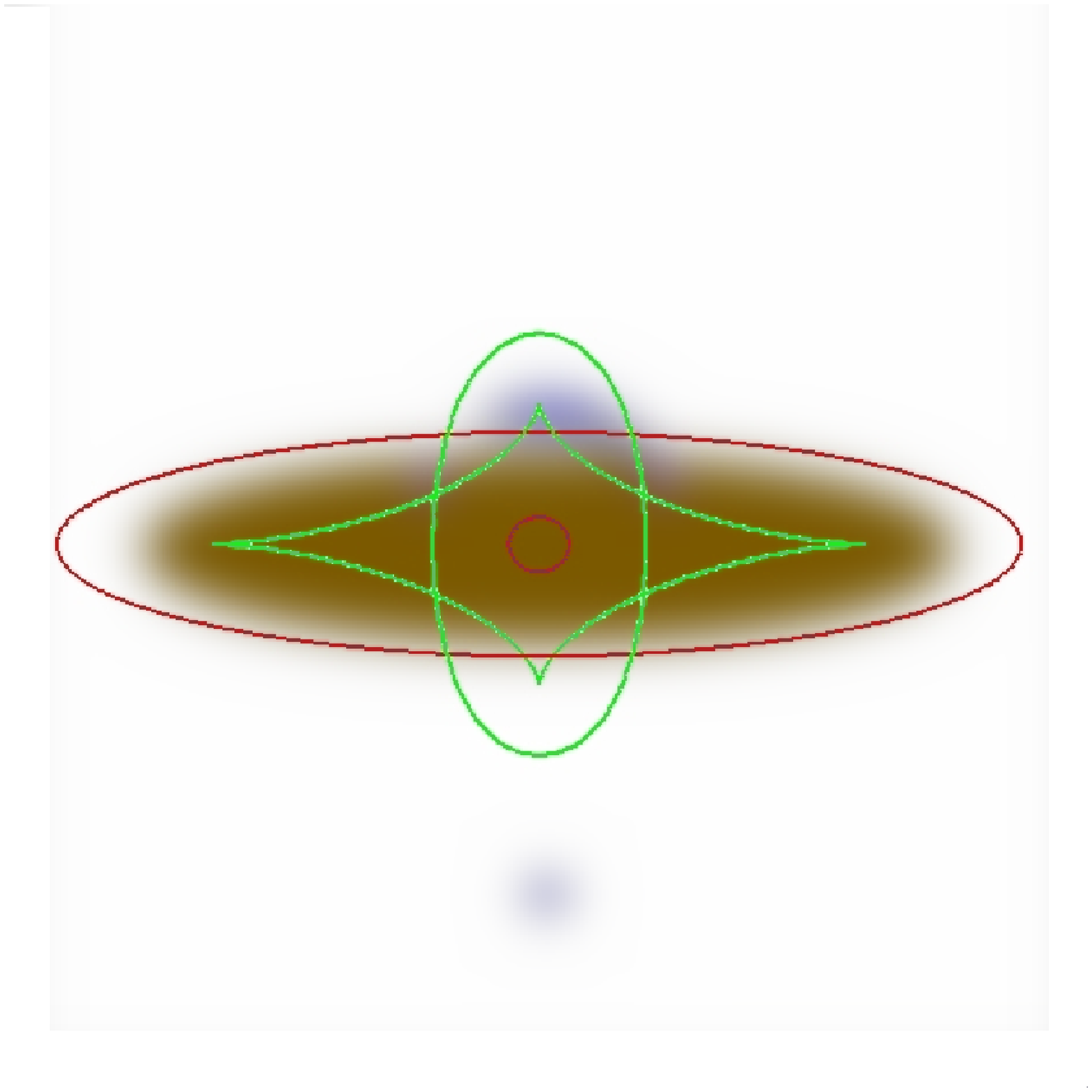}
   \includegraphics[width=4.cm]{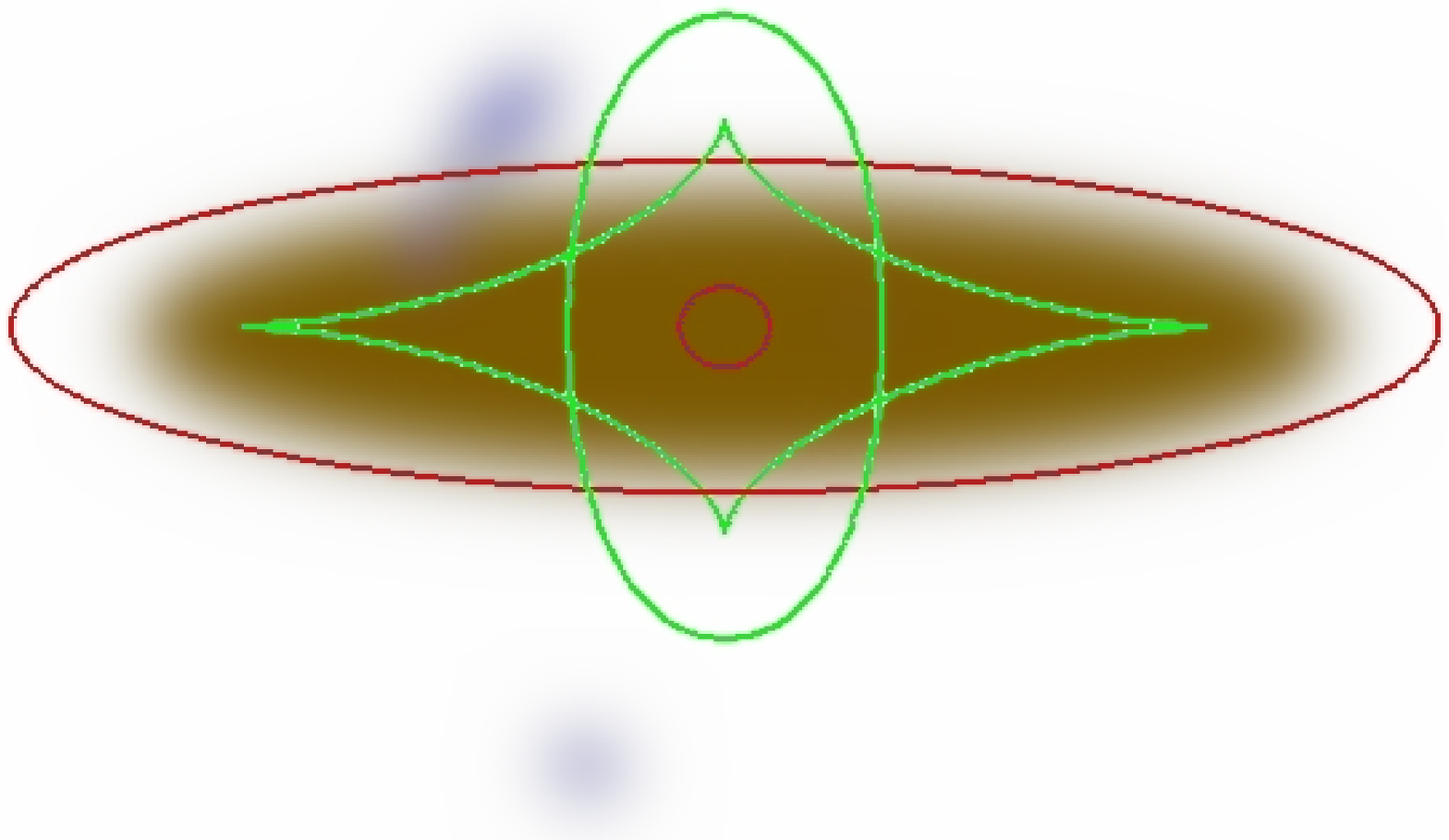}\\
   \includegraphics[width=4.cm]{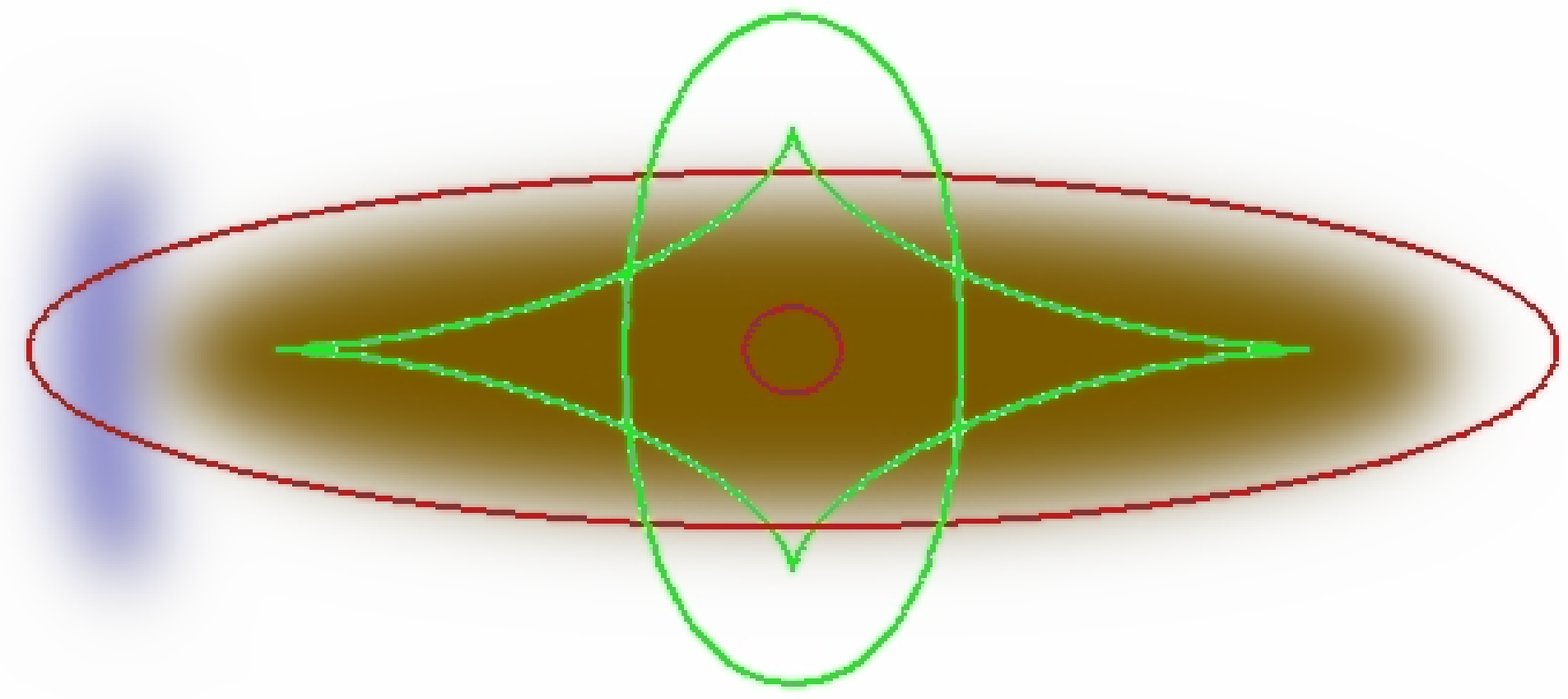}
   \includegraphics[width=4.cm]{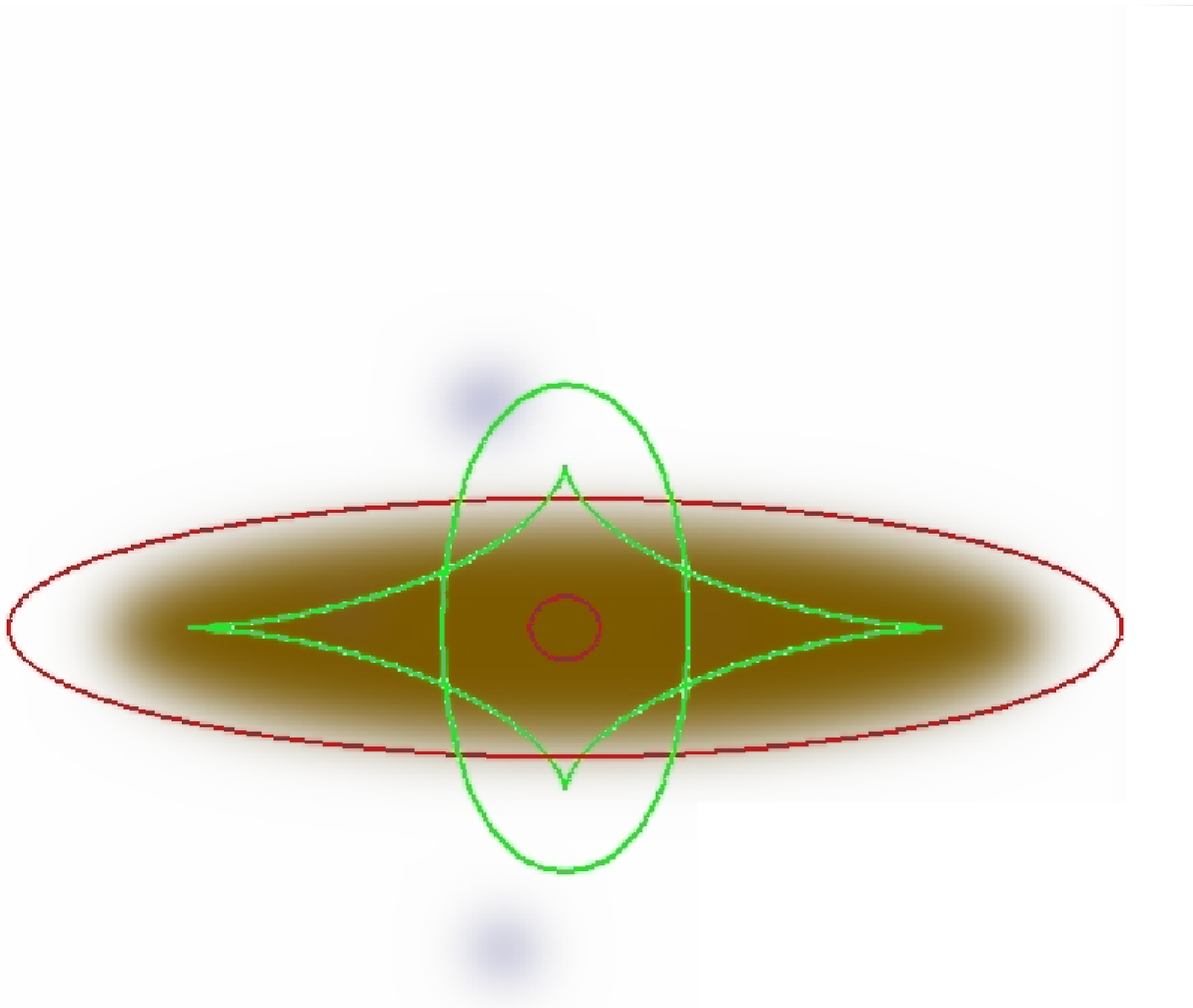}
   \caption{ 
      \finbf{Families of multiple images configurations for lensing 
      by an edge-on disk (yellow/dark) galaxy for different source (blue/light) positions.
      Red/dark lines (resp. green/light) represent critical lines (resp. caustics)
      in the image (resp. source) plane.}  
      \emph{From top left to bottom right:} ``Bulge arcs'', ``Fold arcs'', 
     ``Disk arcs'' and ``Pairs'' \citep{Shin2007}.}
   \label{fig:templates}
\end{figure}

\begin{itemize}
\item[(i)]\emph{Arcs between $0\farcs6$ and $3\farcs5$:} The inner limit is 
  set by the separability of arc and deflector as said above whereas the outer 
  limit is thought to be just above the innermost radius accessible to the automated arc 
  detectors 
  used by the SL2S collaboration for detecting large 
  separation group and cluster scale lenses \citep{Cabanac2007-SL2S}. 
  Note that we discard arcs that are definitely too dim for spectroscopic
  follow-up (estimated to correspond to a magnitude $i\gtrsim24$).
\item[(ii)]\emph{Arc and deflector colors are different:} 
  In order to reduce the contamination by faint satellite galaxies falling 
  into the main galaxy and producing tidal tails, we require the color of 
  the arc and the deflector to be different. 
  This criterium will not be 100\% efficient at removing close pairs
   but its impact on completeness is very low because the probability that a
   lens and a source at very different redshifts have the same color is
   almost negligible.  
   Since precise photometry is difficult on such small scales, color 
   estimation is essentially based on visual inspection of the color images. 
\item[(iii)]\emph{Should look like a library of lensed features:} Finally, 
  the most stringent criterium that we apply is based on the geometry of the 
  arc and its counterimages, as almost all\footnote{The observation of faint 
    counterimages close to the center is very challenging, especially for disk 
    galaxies with substantial dust extinction \citep{Kochanek2006-SAASFE}.} 
  edge-on disk lenses produce very typical image configurations 
  \citep[see e.g.][]{Keeton1998,Bartelmann1998,Shin2007} that we can classify 
  into four families ``Bulge arcs'', ``Fold arcs'', ``Disk arcs'' and ``Pairs'' 
  as suggested by \citet{Shin2007} and illustrated in Fig.~\ref{fig:templates}. 
  Naturally, intermediate cases between these categories may exist with about 
  the same probability for an edge-on disk galaxy similar to the Milky Way 
  \citep{Shin2007}.
\end{itemize}

At this stage we obtain 11 candidates labeled with a A or B upper script  
in Table~\ref{table:long} and also shown in the upper panels panels of 
Fig.~\ref{fig:16candidates}.
\begin{itemize}
\item[(iv)] We then further require that the arc radius is smaller than twice 
	the Einstein radius estimated with the Tully-Fisher method as discussed 
	in \S~\ref{ssec:TF}.
\end{itemize}
This latter filter (hereafter the TF (iv) test) leaves us with only 2 galaxies 
labeled A in Table~\ref{table:long} which would be our prime candidates for 
follow-up observations: they occupy the first row of Fig.~\ref{fig:16candidates}.
We notice that, no matter whether we consider class A or A+B candidates, 
we find numbers that are roughly consistent with our calculation of Sect.~\ref{sec:expected} although we fully acknowledge that completeness cannot be achieved given 
the sharp cutoffs we used.

\subsection{Manual cross-check of the automated procedure} \label{ssec:cross-check}
In order to validate the above procedure and to assess the effects of
photometric redshift uncertainties, we decided to cross-check it by 
visually analyzing the complete sample of \NSP\ objects of the CFHTLS 
selected at the end of \S~\ref{ssec:photometry}. 

The number of objects is large but a trained observer, working in sessions of 
about two hours a day\footnote{This was found to be a decent limit to 
      preserve human eye accuracy. Repeated visual inspections of several 
      fields were done at different periods to assess the stability of the 
      human decision process.}, 
is able to explore about 1\,000 sub-images per hour, corresponding to 4 deg$^2$ per hour. 
This procedure would be impossible for an all-sky survey but it translate into 
about 30 hours for the whole CFHTLS Wide and allows an interesting cross-check 
of the automated procedure. 
 
We obviously apply the same criteria described above in \S~\ref{ssec:visual}.
For information, the first of these steps: (i) \emph{An edge-on galaxy 
must have a faint object nearby}, allows us to get rid of more than 
99.3\% of the potential lenses. 
It leaves us with \PLOS\ interesting galaxies (i.e. $1.5 \pm 0.5$ 
per deg$^2$) that deserve closer inspection, using steps (ii) and (iii).

\begin{figure}
  \label{fig:histogram}
  \centering
  \includegraphics[width=4.3cm]{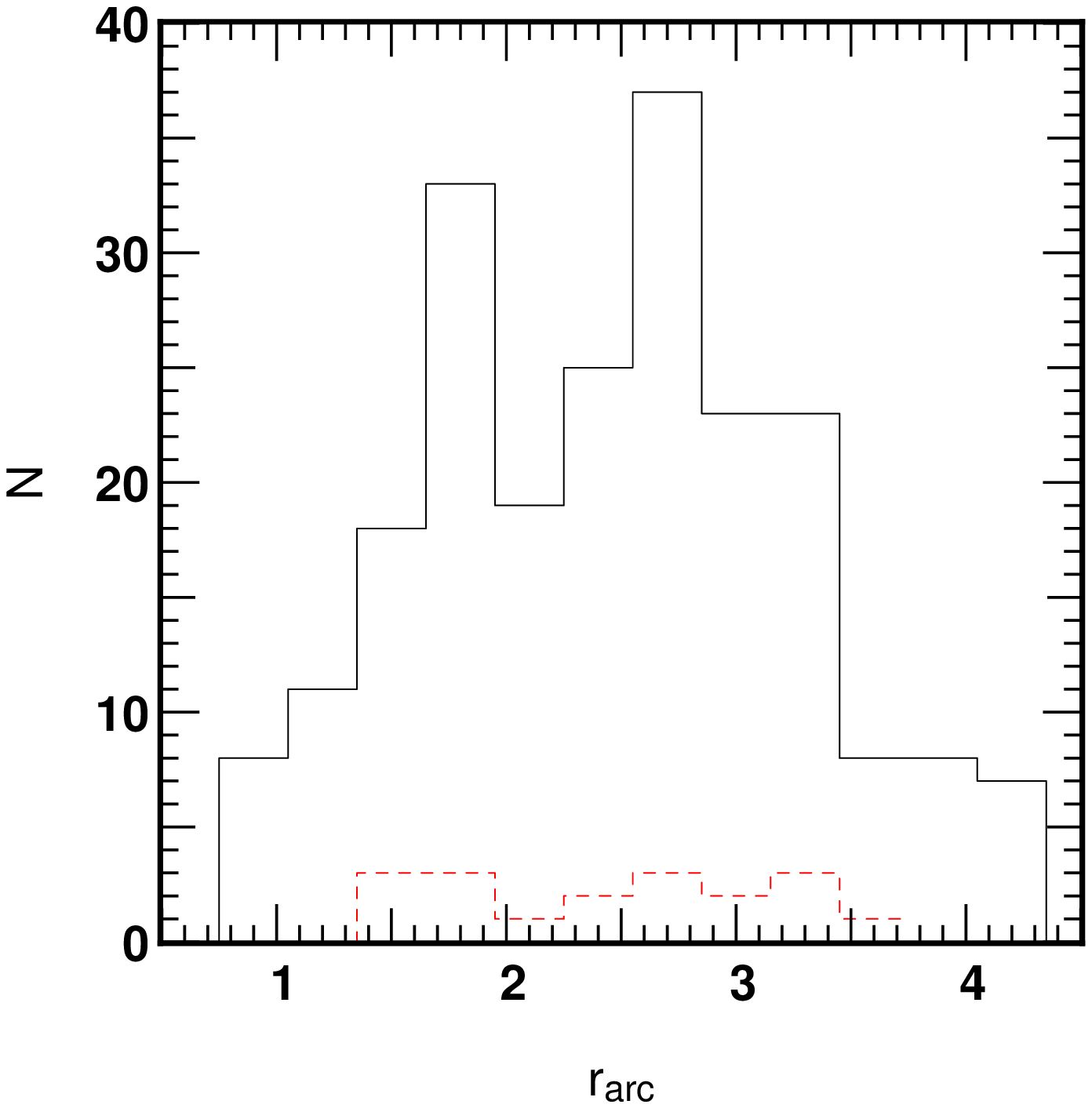}
  \includegraphics[width=4.3cm]{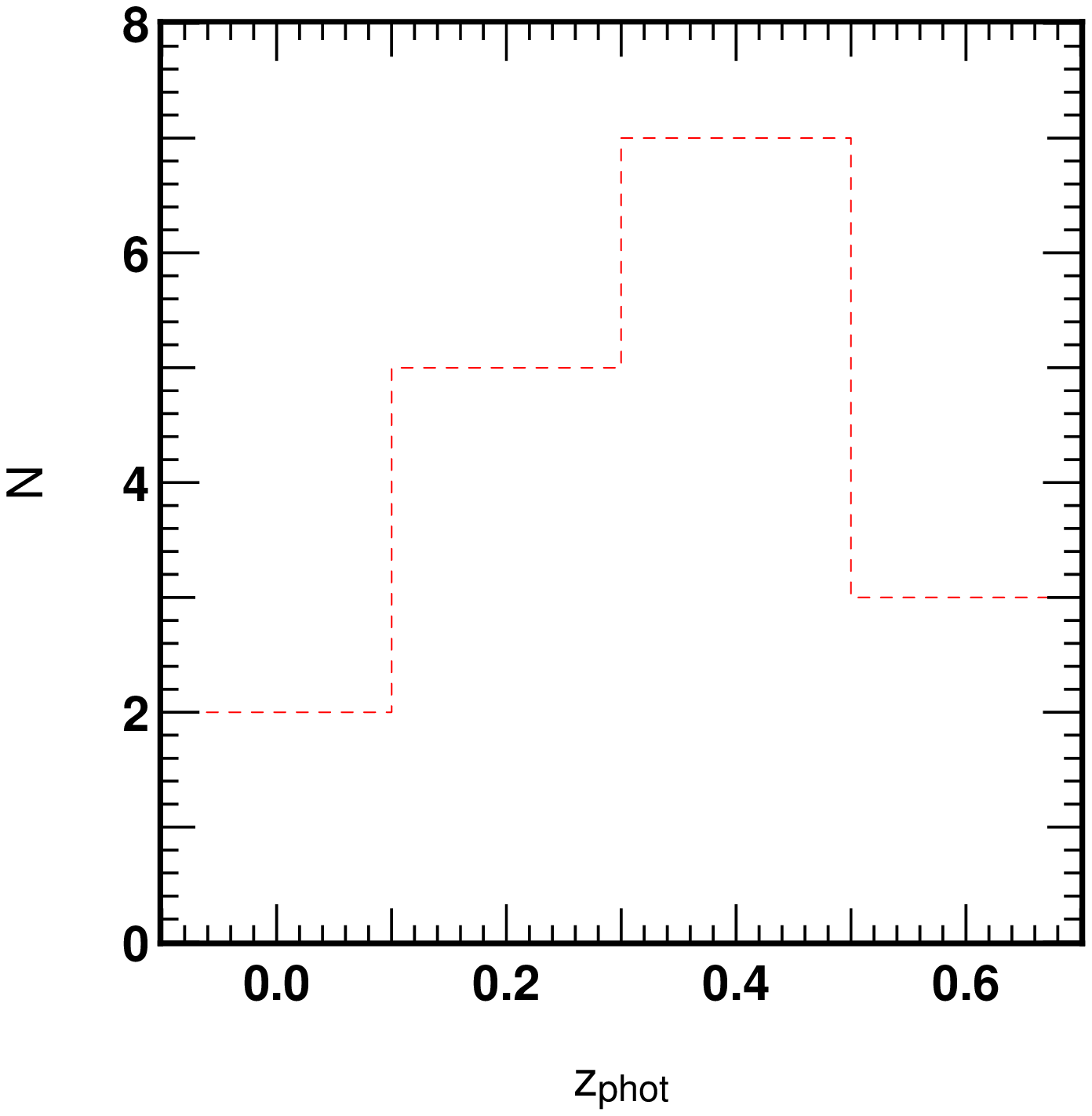}
  \caption{\emph{Left panel:} Distribution of arc radii for 
     the \PLOS\ peculiar lines of sight (solid black curve)
     \finbf{of \S~\ref{ssec:cross-check}}, 
     and for the 16 best candidates (dashed red curve) 
     \finbf{listed in Table~\ref{table:long}.}
     \emph{Right panel:} Photometric redshift distribution for these 16 candidates. 
 }
\end{figure}

\begin{figure*}
   \centering
   \includegraphics[width=4.5cm]{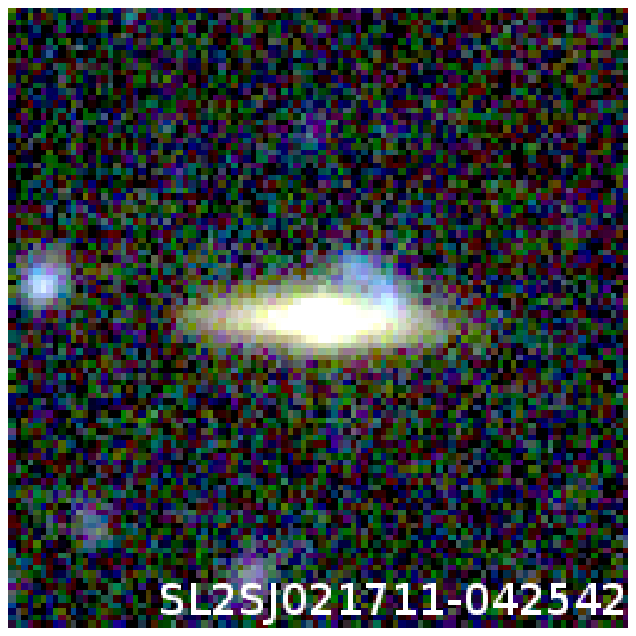} 
   \includegraphics[width=4.5cm]{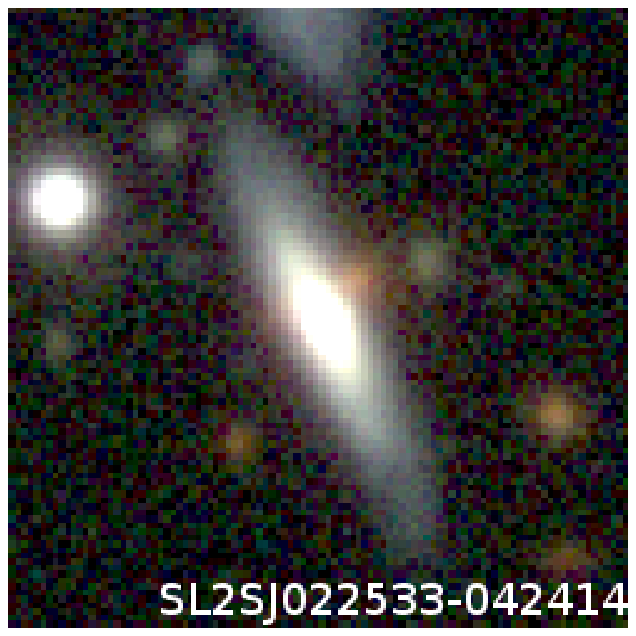} 

   \includegraphics[width=3.0cm]{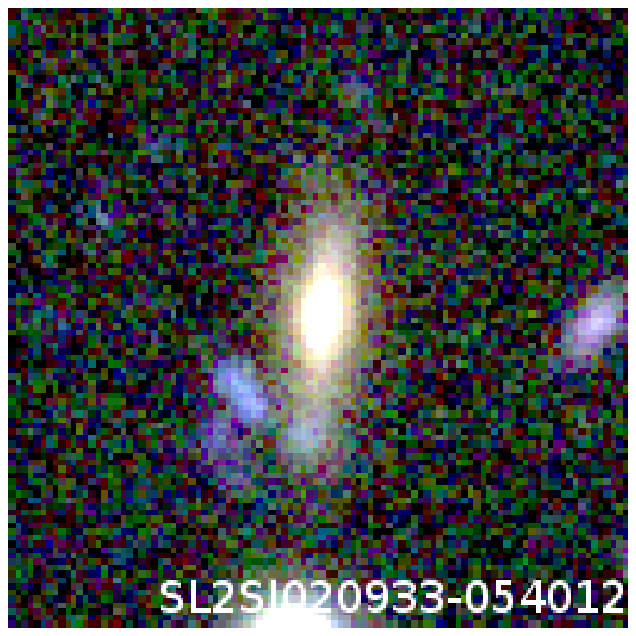} 
   \includegraphics[width=3.0cm]{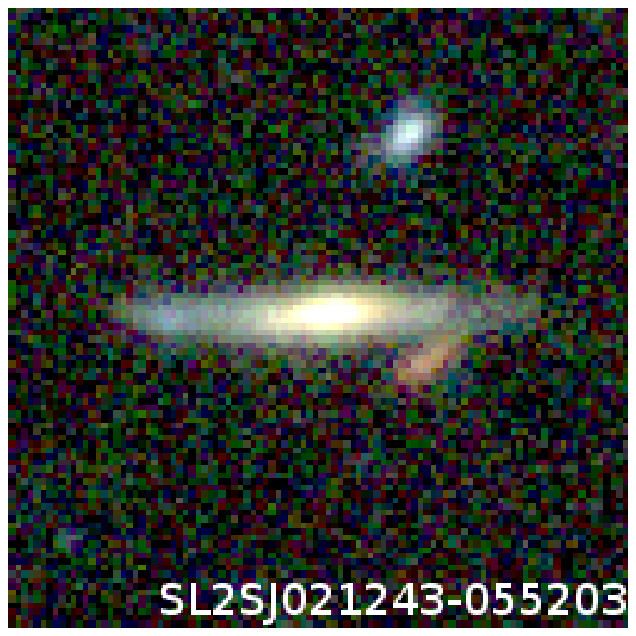} 
   \includegraphics[width=3.0cm]{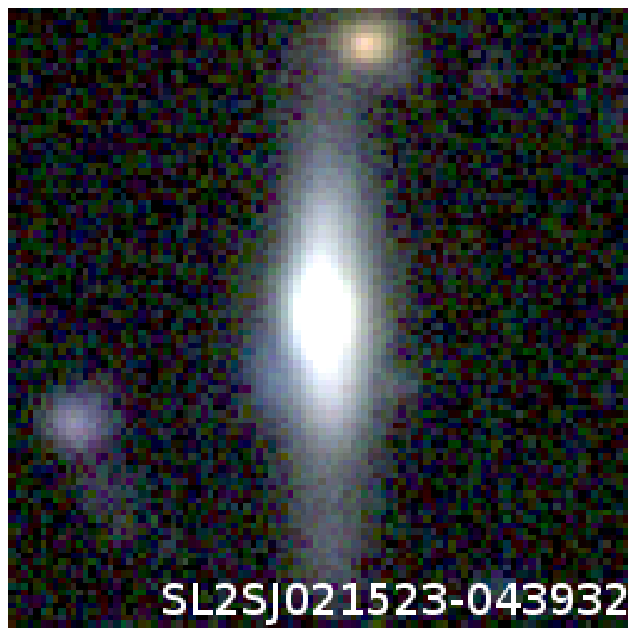} 
   \includegraphics[width=3.0cm]{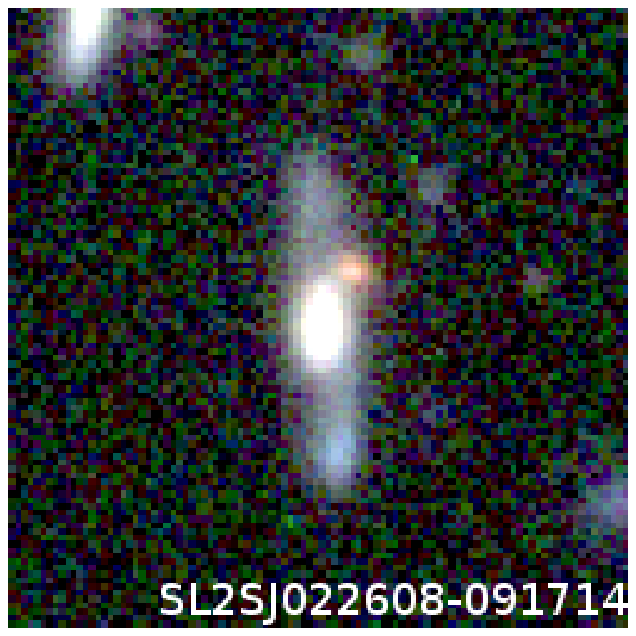} 
   \includegraphics[width=3.0cm]{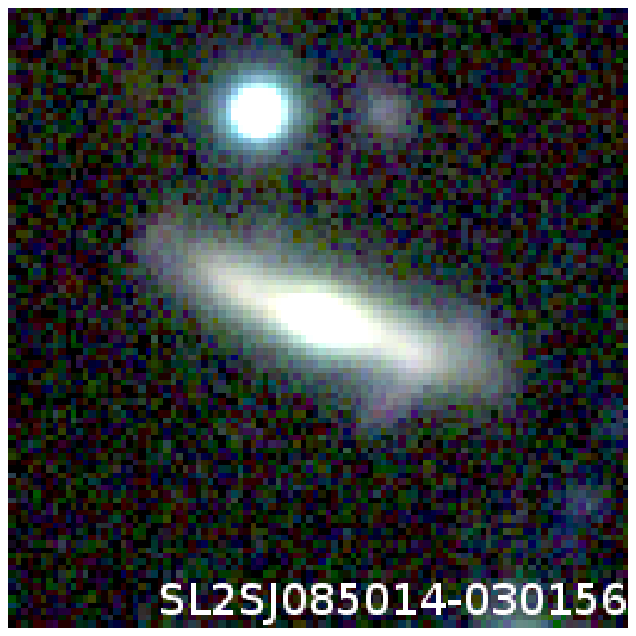} 
   
   \includegraphics[width=3.0cm]{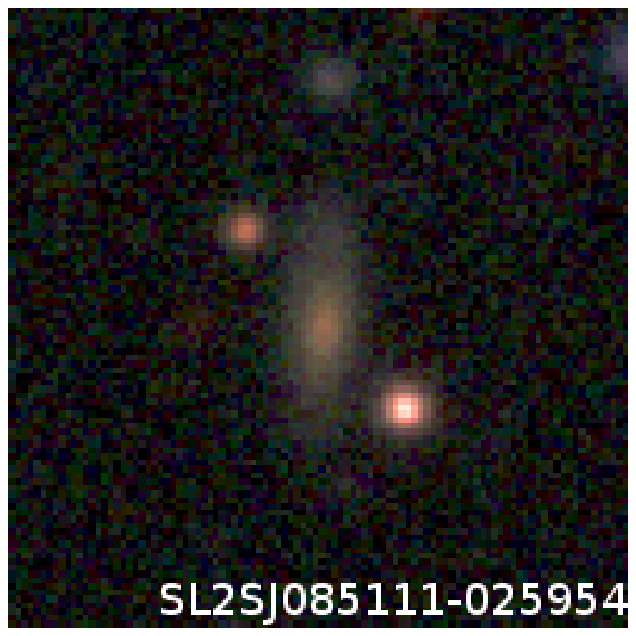} 
   \includegraphics[width=3.0cm]{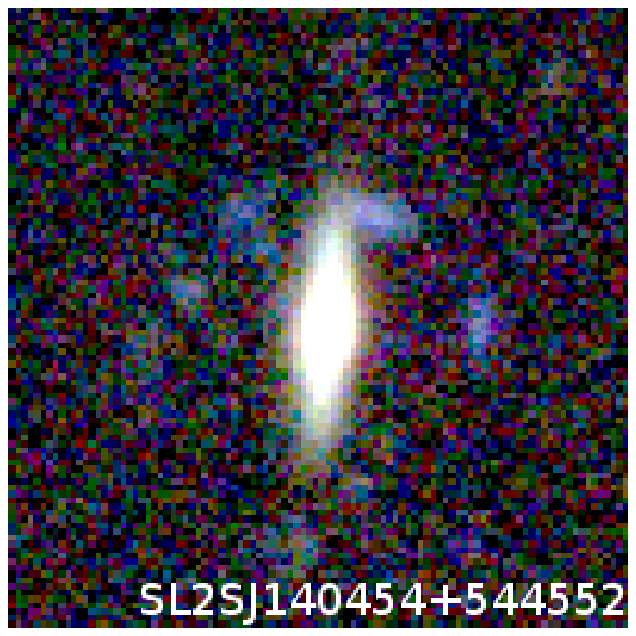} 
   \includegraphics[width=3.0cm]{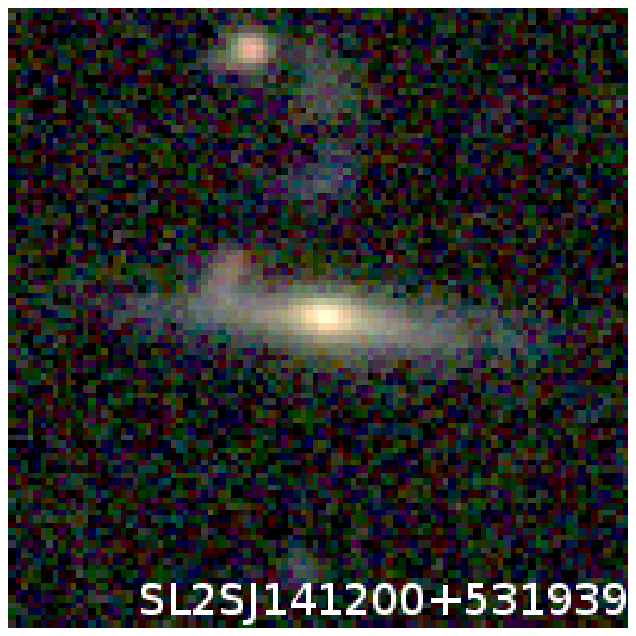} 
   \includegraphics[width=3.0cm]{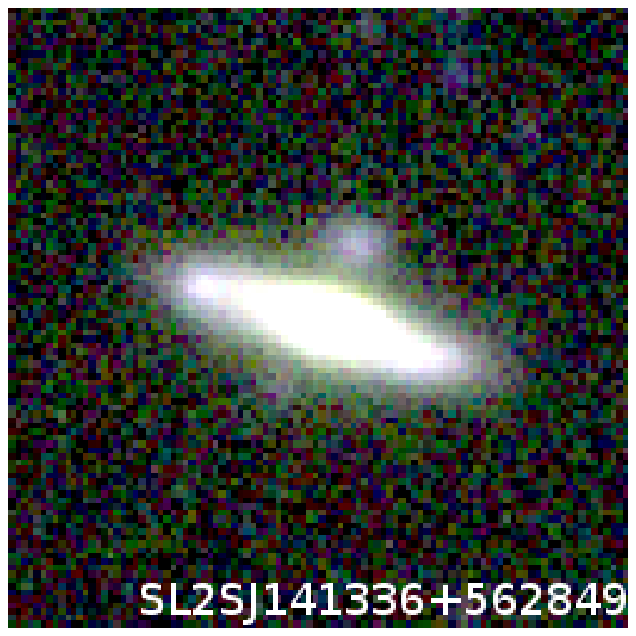} 
   
   \includegraphics[width=3.0cm]{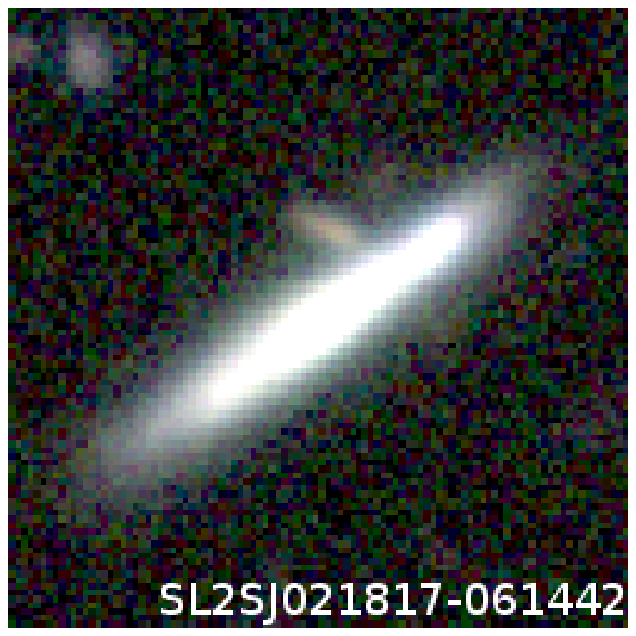} 
   \includegraphics[width=3.0cm]{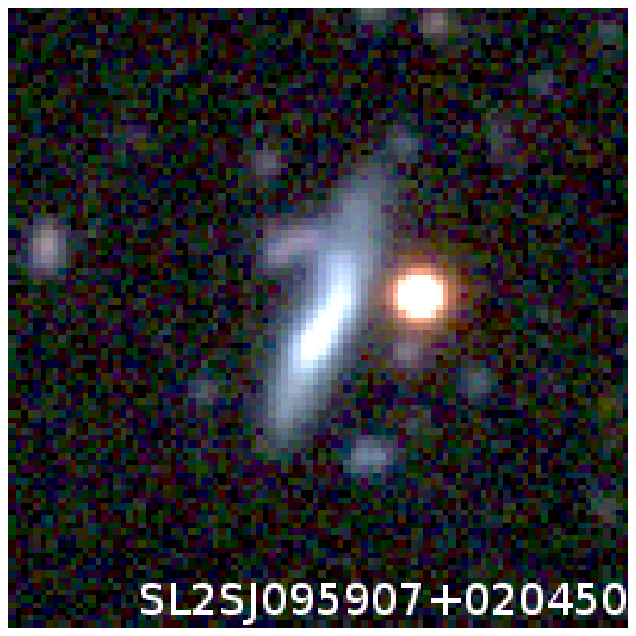} 
   \includegraphics[width=3.0cm]{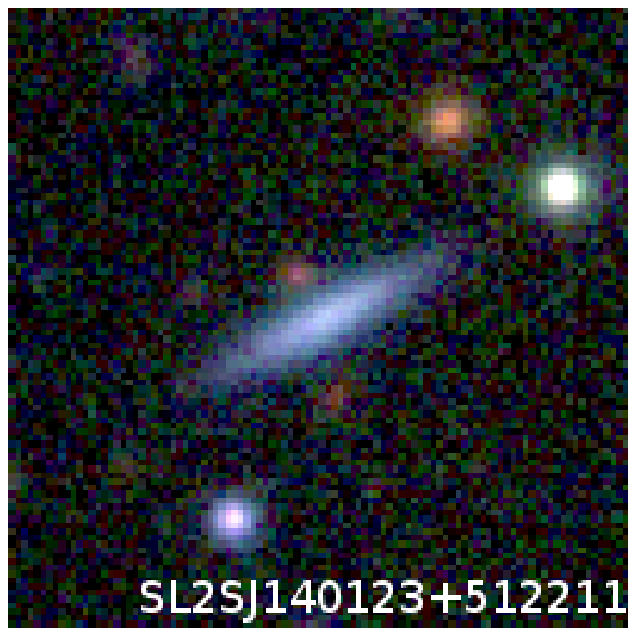} 
   \includegraphics[width=3.0cm]{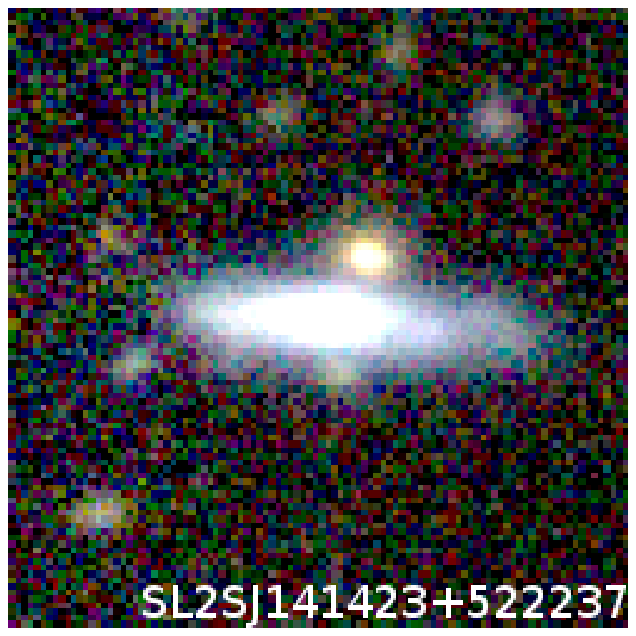} 
   \includegraphics[width=3.0cm]{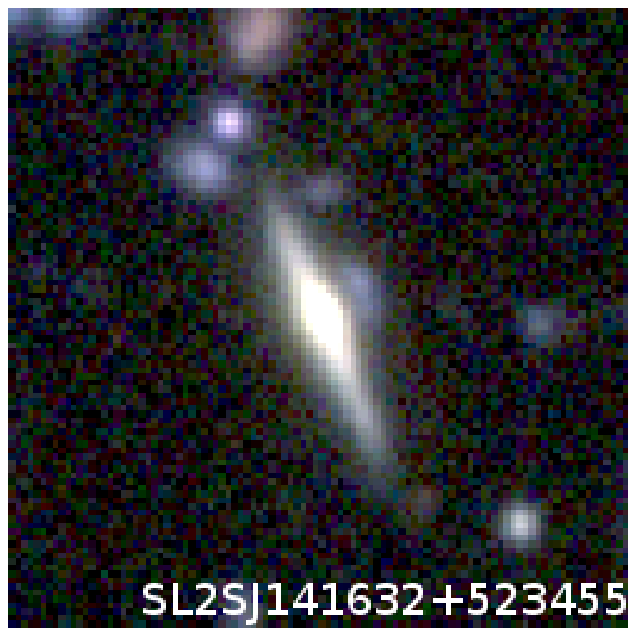} 

   \caption{Color images of the 16 candidates. Image sizes are all 
      $18\farcs 6$ on a side.
      \emph{First row:} Class A objects passing the Tully-Fischer (iv) test.
      \emph{Second and third rows:} Good B quality systems that do not pass the 
         TF (iv) test. 
      \emph{Fourth row:} Additional \finbf{class} C systems found during manual cross-check.}
   \label{fig:16candidates}
\end{figure*}

The full visual selection procedure allowed to keep only 16 candidates. 
Of them, 11 were already selected as class A or B in the more automated procedure above whereas 5 additional systems were found to be worth presenting here although they don't pass the Tully-Fisher tests (neither the first cut $2 \reTF \ge 0\farcs8$ nor the second TF (iv) 
test $\rarc \le 2 \reTF$). 
These class C objects occupy the bottom of Table~\ref{table:long} and Fig.~\ref{fig:16candidates}, while Fig.~\ref{fig:histogram} shows the photometric redshift distribution and the distribution of arc radii for the full 16 candidates.

This time-consuming cross-check therefore validates our use of 
the TF tests, and shows that errors on photometric redshifts are not, 
in our case, a major source of concern.

\begin{table*}[htbp]
   \caption{Edge-on disk lens candidates found in the CFHTLS Wide survey.}
   \label{table:long}
   \begin{tabular}{lccccccccccc}
   \hline\hline
Name & RA (J2000)   & DEC (J2000)    & \multicolumn{5}{c}{AB mag}
     & $\rarc$      & $\zl$ & Family & $\reTF$     \\
     & (hr~min~sec) & ($\degr$~'~")  & $u^*$ & $g$ & $r$ & $i$ &$z$  
     & $(\arcsec)$  &       &        & ($\arcsec$) \\
    \hline

SL2SJ021711$-$042542\tablefootmark{A} & 02:17:11.843 & -04:25:42.00   &
22.80 & 21.60 & 20.43 & 19.72 & 19.35 & 1.7 & 0.57$\pm$0.06 & \cB     & 0.90 \\

SL2SJ022533$-$042414\tablefootmark{A} & 02:25:33.129 & -04:24:14.99   &
20.79 & 19.50 & 18.68 & 18.18 & 17.85 & 1.4 & 0.24$\pm$0.04 & \cB/\cD & 0.70 \\

\hline

SL2SJ020933$-$054012\tablefootmark{B} & 02:09:33.633 & -05:40:12.99   &
23.12 & 21.75 & 20.39 & 19.75 & 19.42 & 3.3 & 0.44$\pm$0.04 & \cF     & 0.61 \\

SL2SJ021243$-$055203\tablefootmark{B} & 02:12:43.801 & -05:52:03.23   &
22.68 & 21.38 & 20.29 & 19.76 & 19.44 & 3.0 & 0.35$\pm$0.06 & \cF/\cD & 0.44 \\

SL2SJ021523$-$043932\tablefootmark{B} & 02:15:23.801 & -04:39:32.33   &
21.14 & 19.64 & 18.76 & 18.18 & 17.89 & 2.2 & 0.21$\pm$0.04 & \cB/\cF & 0.58 \\

SL2SJ022608$-$091714\tablefootmark{B} & 02:26:08.138 & -09:17:14.87   &
22.35 & 21.16 & 20.11 & 19.52 & 19.11 & 1.8 & 0.46$\pm$0.06 & \cF     & 0.76 \\

SL2SJ085014$-$030156\tablefootmark{B} & 08:50:14.572 & -03:01:56.08   &
21.73 & 20.35 & 19.21 & 18.57 & 18.17 & 3.1 & 0.44$\pm$0.04 & \cD     & 1.28 \\ 

SL2SJ085111$-$025954\tablefootmark{B} & 08:51:11.743 & -02:59:54.21   &
23.79 & 22.71 & 21.49 & 20.74 & 20.23 & 3.3 & 0.55$\pm$0.04 & \cP     & 0.44 \\

SL2SJ140454$+$544552\tablefootmark{B} & 14:04:54.217 & +54:45:52.49   &
22.79 & 21.17 & 19.92 & 19.30 & 18.86 & 3.6 & 0.41$\pm$0.04 & \cD     & 0.74 \\

SL2SJ141200$+$531939\tablefootmark{B} & 14:12:00.934 & +53:19:39.83   &
22.38 & 21.33 & 20.35 & 19.81 & 19.55 & 3.2 & 0.42$\pm$0.06 & \cD     & 0.55 \\

SL2SJ141336$+$562849\tablefootmark{B} & 14:13:36.001 & +56:28:49.25   &
21.39 & 20.01 & 19.11 & 18.60 & 18.26 & 2.3 & 0.27$\pm$0.04 & \cB     & 0.63 \\

\hline

SL2SJ021817$-$061442\tablefootmark{C} & 02:18:17.050 & -06:14:42.59  &           
20.13 & 18.93 & 18.44 & 18.04 & 17.91 & 2.8 & 0.07$\pm$0.06 & \cD    & 0.13 \\  

SL2SJ095907$+$020450\tablefootmark{C} & 09:59:07.769 & +02:04:50.30  & 
22.22 & 21.46 & 21.25 & 20.97 & 20.84 & 2.8 & 0.07$\pm$0.06 & \cF    & 0.22 \\ 

SL2SJ140123$+$512211\tablefootmark{C} & 14:01:23.392 & +51:22:11.00  &
22.24 & 21.22 & 20.86 & 20.47 & 20.31 & 1.4 & 0.14$\pm$0.10 & \cP    & 0.08 \\

SL2SJ141423$+$522237\tablefootmark{C} & 14:14:23.150 & +52:22:37.46  &
21.02 & 19.94 & 19.48 & 19.27 & 19.08 & 1.9 & 0.17$\pm$0.04 &  \cP   & 0.22 \\

SL2SJ141632$+$523455\tablefootmark{C} & 14:16:32.336 & +52:34:55.14  & 
22.86 & 21.62 & 20.78 & 20.35 & 19.96 & 1.6 & 0.30$\pm$0.06 &  \cB   & 0.25 \\ 

\hline
\end{tabular} 
\tablefoot{
 \finbf{ The uncertainty on magnitudes is, in almost every case, of the 
 order of 0.01 and the Einstein radius is the maximum allowed by the TF law. 
 For geometry family, \cB~stands for ``bulge arc'', \cF~for ``fold arc'', 
 \cD~for ``disk arc'' and \cP~for ``pair'' (see \S~\ref{ssec:visual}).\\
 \tablefoottext{A}{Candidates found with both the automated and the manual 
 cross-check procedures and satisfying the Tully-Fisher (iv) test.}
 \tablefoottext{B}{Candidates same as $^{(A)}$, but failing the Tully-Fisher (iv) test.}
 \tablefoottext{C}{Extra candidates found only with the manual cross-check procedure.} 
 }
 }
\end{table*}

\section{Results \& Discussion} \label{sec:discussion}

\subsection{Description of the candidates}
We have seen that our search of edge-on disk galaxies in 124~deg$^2$ of the CFHTLS Wide yields 2 A-class good candidates (listed at the top of Table~\ref{table:long} and Fig.~\ref{fig:16candidates}). We believe they are the only genuine strong lenses detected by our procedure.
\begin{itemize}
\item SL2SJ021711-042542 is a spiral galaxy with a photometric redshift
 	 $ \zl = 0.57 \pm 0.06$. It is a bulge arc configuration with an observed arc radius ($\rarc =1\farcs7$) fully compatible with the maximum arc radius  allowed by a TF model ($2 \reTF = 1\farcs8$).
 \item SL2SJ022533-042414 is also a spiral with a photometric redshift $\zl = 0.24 \pm 0.04$. It is a bulge arc configuration at $\rarc \simeq 1\farcs4$ with a faint red counter image which is compatible with the maximum arc radius of the TF model ($2 \reTF = 1\farcs4$).
\end{itemize}

From a pure visual selection (i.e. not applying filter (iv)),
we would have selected 9 more candidates (labeled B in Table~\ref{table:long}).
These uncertain candidates can be roughly classified in two categories. 
\begin{itemize}
\item 85\% of them have an arc radius that is too large for their lens 
magnitude and a small color difference with their lens. 
They probably correspond to stellar tidal trails 
produced by galaxy interactions or disrupted satellites.
For the most massive satellites falling in the centre of a galaxy potential 
these structures can be bright enough to be observed, especially when star 
formation is initiated by gravitational interactions. 
These thin structures are wrapped around the dark matter halo potential 
and can mimic arc structures.
Some of them are seen straddling the edge-on disk like an arc triplet 
formed by a naked cusp. 
We see in Fig.~\ref{fig:histogram}'s left panel a cut-off for $\rarc > 2\farcs5$. 
Its existence shows that this main source of spurious detections are more easily 
discriminated for large arc radii.
Only the high spatial resolution of HST images could discriminate them at smaller 
radius.
\item The remaining 15\% (namely SL2SJ020933-054012 and SL2SJ140454+544552) 
correspond probably to Singly Highly Magnified Sources (SHMS) where the 
arc is formed at a distance greater than $ 2 \re $ and does not produce multiple images. 
The optical depth of this strong flexion regime was studied by 
\citet{Keeton2005} with both isothermal and NFW halo density profiles.
He found that the proportion of SHMS among arcs should be less than $20\%$ for 
distant background sources. 
For these candidates we shall remember that the baryonic mass within the disk can 
increase the convergence of the edge-on lens beyond the one considered for the TF 
test \citep{Bartelmann1998}. 
Therefore if the arc is not too far from the critical line an observer shall not 
systematically discard them without further considerations.
\end{itemize}

We also listed 5 class C candidates at the bottom of Table~\ref{table:long} 
which have been selected only during the cross-check manual procedure 
(\S~\ref{ssec:cross-check}), because their theoretical Tully-Fisher 
radius $\reTF $ is less than half the seeing value. 
We believe they are interesting systems anyway and are worth presenting here.

\subsection{Discussion \& Prospects}
During this detection exercise, several lessons were found, which can be 
of some use for future surveys.

The Tully-Fisher cut turns out to be an efficient way 
to strongly decrease the number of line of sight to be visually scrutinized
and to get rid of many of the false positive cases. 
Indeed no bona fide strong lens was found during the manual cross-check 
(\S~\ref{ssec:cross-check}) out of the $\NSP-\NTF$ objects discarded by 
the TF cut (\S~\ref{ssec:TF}).
However we stress that the accuracy of this cut is limited by the 
quality of photometric redshifts. 
With a limited number of colors, high redshifts could be mistaken for 
low ones, specially in the range $ 0.05 < \zl < 0.15 $ 
\citep[e.g.][]{Bernstein2009}, therefore the 
absolute magnitude of the lens would be over estimated and thus resulting in too small a value of $\reTF$. We therefore recommend not to implement a too stringent TF cut.

We show in the left panel of Fig.~\ref{fig:histogram} the distribution of arc radii.
The effect of the seeing cut-off at small radii can be clearly seen:
the numerous lenses with arc radius smaller than the seeing disk 
($0\farcs8$) are missed. 
The photometric redshift distribution of the 16 selected candidates 
lies in the range $0.2 \lesssim \zl \lesssim 0.6$ as 
shown in the right panel of Fig.~\ref{fig:histogram}. 
In this respect, our CFHTLS candidates nicely complement the ones extracted from
shallower surveys like SDSS \citep[see e.g.][]{Feron2009}.

We can wonder if the LSST \citep{LSST2009} survey will significantly improve the situation. 
It will cover 20\,000 square degrees of the sky and will reach the depth 
of the CFHTLS Wide just after the first year of observation and 
the depth of the CFHTLS Deep after ten years \citep{Ivezic2008}.
The seeing quality at Cerro Pachon in Chile is expected to be $0\farcs7$ 
quite comparable to the one at CFHT. Therefore, in theory, we can expect from our 
calculation (cf. Fig.~\ref{fig:nlens}) at least 1\,300 massive edge-on lenses in LSST.  
Its deeper photometry will also allow to get better photometric redshift and 
hence to improve the chance of keeping good candidates with the TF test. 
Nevertheless, conservatively assuming the same detection probability as in CFHTLS,
one expects to find in the LSST about $1770$ candidates instead of the $11$ A+B 
class candidates found here, leading probably to about 320 bona-fide edge-on lenses,
which is an excellent perspective.

The situation shall improve even more dramatically with future space borne surveys
like JDEM or Euclid
\citep{Marshall2005, Refregier2009} which will have a better resolution ($< 0\farcs3$).
We expect from our calculation that these surveys could detect a few thousand 
edge-on lenses with Einstein radius substantially smaller than in ground based surveys.

\section{Conclusion}
In conclusion, while it is extremely difficult to find edge-on lenses with ground based 
imaging surveys, the situation is not completely \finbf{hopeless} with sub-arcsecond 
seeing ($< 0\farcs8$). 

A good way to implement a fast search procedure is to preselect 
a limited subset of galaxies presumably massive enough 
to produce a detectable arc. 
The Tully-Fisher law seems quite appropriate for this purpose, 
since it can predict an Einstein radius for each disk.
Galaxies with the largest Einstein radii are 
then checked for the presence of a faint source nearby
which might be a gravitational arc.
Extensive visual cross-checking of 
all disk galaxies present in the survey has shown that this procedure 
picks most of the interesting candidates, provided reliable photometric
redshifts are available.

Whatever the final choice of an observer, the number of candidates that 
can be choosen for an observational follow up is quite small: 
here we ended up with only 2 good candidates from 124 square degrees 
of the CFHTLS Wide. 
This illustrates how challenging it is to dig out such lenses from any 
ground based survey. 
Despite their great number in the sky and as long as wide space surveys 
are not available, edge-on disk lenses will remain scarce and precious 
objects for astronomers who try to understand
the relative distribution of dark and ordinary matter in spirals.

\begin{acknowledgements} 
The authors are thankful to the CFHTLS members and the Terapix team for 
their excellent work in reducing and distributing data to the community.
They would also like to thank P. Marshall and J.-P. Kneib for fruitful 
discussions, and J. Coupon for providing his photometric redshifts calculations.
Part of this work was supported by the Agence Nationale de la Recherche (ANR)
and the Centre National des Etudes Spatiales (CNES). 
Part of this project is done under the support of the National Natural Science
Foundation of China No.10878003, 10778752, Shanghai Foundation No. 07dz22020, 
and the Leading Academic Discipline Project of Shanghai Normal
University (08DZL805).
\end{acknowledgements}

\bibliographystyle{aa}      
\bibliography{3977sygnet}

\end{document}